\documentclass[aps,prd,reprint,amsmath,amssymb,nofootinbib,superscriptaddress]{revtex4-1}
\usepackage[T1]{fontenc}
\usepackage{lmodern,graphicx,bm,booktabs}
\usepackage[colorlinks=true,linkcolor=blue,citecolor=blue,urlcolor=blue]{hyperref}
\newcommand{\dd}{\mathrm{d}}
\newcommand{\Tc}{T_{\mathrm c}}
\newcommand{\Bbag}{\mathcal B}
\newcommand{\cE}{\mathcal E}
\newcommand{\cB}{\mathcal H}
\newcommand{\OO}{\mathcal O}
\begin{document}
\title{Deconfinement in weak electric and magnetic fields}
\author{Eduardo S. Fraga}
\affiliation{Instituto de F\'isica, Universidade Federal do Rio de Janeiro,
Caixa Postal 68528, Rio de Janeiro, RJ 21941-972, Brazil}
\author{Let\'icia F. Palhares}
\affiliation{Departamento de F\'\i sica Te\'orica,
Universidade do Estado do Rio de Janeiro,
Rua S\~ao Francisco Xavier 524, 20550-013 Maracan\~a,
Rio de Janeiro, Brazil}

\begin{abstract}

We investigate the thermal deconfinement transition in weak electric and magnetic
fields, describing the hadronic phase by a pion gas and the quark--gluon
plasma by the MIT bag model. Besides the thermal response of each phase, we
include the one-loop electromagnetic vacuum terms of quarks and charged pions,
renormalized at a common scale that fixes the charge in the confined vacuum. With no fitting of the electromagnetic response, the critical temperature rises with the electric field and falls with the magnetic one. The electric curvature is of the same size as that obtained on the lattice, while the magnetic one is a few times larger. We discuss the dependence on quark mass and renormalization scale, and a large-$N_c$
organization of both curvatures, whose sum is independent of the vacuum terms
and connects directly to perturbative QCD.

\end{abstract}
\maketitle

\section{Introduction}
\label{sec:intro}

The thermodynamics of strong interactions in the presence of external electromagnetic fields has attracted increasing interest for decades. Phenomenologically, they are relevant in a variety of scenarios \cite{Adhikari:2024bfa}, from the astrophysics of compact stars \cite{Duncan:1992hi,Thompson:1993hn,Kouveliotou:1998ze} and binary neutron star mergers \cite{Jiang:2025ijp,Most:2025kqf} to non-central, high-energy heavy ion collisions \cite{Kharzeev:2007jp,Skokov:2009qp,Voronyuk:2011jd,Bzdak:2011yy,Deng:2012pc,Inghirami:2016iru,Roy:2017yvg}, as well as in the early universe \cite{Vachaspati:1991nm,Enqvist:1993np,Grasso:2000wj}. 
First-principle calculations in QCD can be performed using lattice simulations \cite{DElia:2010abb,Bali:2011qj,Ilgenfritz:2012fw,Bali:2012zg,Bornyakov:2013eya,Bali:2013esa,Bruckmann:2013oba,Bali:2014kia,Endrodi:2015oba,DElia:2018xwo,DElia:2021tfb,DElia:2021yvk,Astrakhantsev:2024mat} (see Ref. \cite{Endrodi:2024cqn} for a recent review), perturbation theory \cite{Blaizot:2012sd,Ayala:2014uua,Ayala:2015bgv,Fraga:2023cef,Fraga:2023lzn,Fraga:2024klm,Endrodi:2022wym,Ferreira:2023cqw,Ferreira:2026qqu}, hard thermal loop perturbation theory \cite{Rath:2017fdv,Haque:2017nxq,Karmakar:2019tdp,Bandyopadhyay:2017cle,Karmakar:2020mnj}, and in certain limits of QCD such as for a large number of colors $N_c$ \cite{Fraga:2012ev} and within chiral perturbation theory \cite{Colucci:2013zoa,Hofmann:2020dvz,Hofmann:2020ism,Adhikari:2021bou}. Of course, one can also adopt holographic models \cite{Preis:2010cq,Preis:2011sp,Preis:2012fh,Finazzo:2016mhm,Critelli:2016cvq} and a variety of effective models \cite{Fraga:2012rr,Kharzeev:2013jha,Andersen:2014xxa,Miransky:2015ava,Andersen:2021lnk,Adhikari:2024bfa}.

Such fields provide additional control parameters to probe the phase
structure of QCD, and lattice simulations have started to map it in some
detail. With physical quark masses, the pseudocritical temperature was found
to decrease in a magnetic background~\cite{Bali:2011qj}, which was quite a surprise given the first predictions within effective models \cite{Fraga:2008qn,Mizher:2010zb,Gatto:2010pt} and preliminary lattice results at the time \cite{DElia:2010abb} (for a review, see Ref. \cite{Adhikari:2024bfa}). For weak electric fields,
on the other hand, the response of the Polyakov loop indicates that the
transition temperature increases, according to recent lattice simulations~\cite{Endrodi:2023wwf}. These opposite trends raise a
simple question: which features of the confined and deconfined phases are
sufficient to account for their signs?

The electric case, however, requires some care before such a comparison can
be made. At nonzero temperature the static electric response depends on the
equilibrium prescription adopted, and perturbative and lattice constructions
based on spatial averaging in a canonical ensemble need not agree with a
homogeneous Schwinger effective action~\cite{Endrodi:2022wym,Carignano:2017ovz,Ferreira:2023cqw,Endrodi:2026kmb}. Throughout this
paper we adopt the former prescription, since we compare our results with observables that can be computed on the lattice. The observables discussed below, therefore, describe a reduced equilibrium potential, with the macroscopic contribution from charge redistribution separated out; they do not, of course, describe the real-time evolution of an accelerating plasma. At nonzero temperature both phases contain charged particles and screen a static field. Within the canonical prescription this screening is accounted for by the macroscopic charge redistribution that has been separated out, and the coexistence condition used below compares the reduced potentials of the two phases at the same value of the external field\footnote{ Whether the electric field ($E$ or $D$) (respectively $H$ or $B$, in the magnetic case) is held fixed affects the pressure balance only at relative order
$e^2|\Delta\xi|,\,e^2|\Delta\chi|\lesssim 10^{-2}$, which we neglect. Here, $e$ is the fundamental electric charge, and $\xi$ and $\chi$ are the electric and magnetic susceptibilities, respectively.}. Notice also that this two-phase construction is not identical to the single-phase lattice criterion of Ref.~\cite{Endrodi:2023wwf}.

In this paper we address the behavior of the deconfinement temperature in
weak electric and magnetic fields within the same economical approach that
was successfully used for the magnetic case under ultra high fields in Ref.~\cite{Fraga:2012fs}, namely a gas of massive pions for the hadronic sector and the MIT bag pressure for the quark--gluon plasma. Even though this description is known to be crude in numerical precision and misses the correct (crossover) nature of the transition, it provides a simple setup to identify the mechanism behind the signs of the shifts in the critical temperature, which is our main purpose. So, while in Ref.~\cite{Fraga:2012fs} we considered very strong magnetic fields, here we work to quadratic order in the fields and do not attempt to describe the saturation of the critical temperature at large fields.

We find that, for light quarks, the electric field penalizes the deconfined
phase more strongly than the pion gas, whereas the magnetic field favors it.
As a result, the shifts in the transition temperature have the same signs
as found on the lattice, {\it without any fitted field dependence of the bag constant or of a coupling}. A crucial ingredient, as in the magnetic case~\cite{Fraga:2012fs}, is the treatment of vacuum contributions. Since the charged particles are different on the two sides of the transition, their one-
loop electromagnetic vacuum terms do not cancel in the pressure balance, and a common renormalization of the electric charge leaves a finite relative term. We compute it with both phases renormalized at a fixed scale, chosen so that $e$ is the charge of the confined vacuum, as also done on the lattice. This relative term reduces both curvatures by
a factor of about four with respect to a purely thermal
treatment, and brings the electric curvature to the size
found in simulations. We also discuss the dependence of
the results on the quark masses and on the choice of
renormalization scale.

We also extend the large-$N_c$ pressure-balance argument of
Ref.~\cite{Fraga:2012ev} to the static electric response. At fixed number of
flavors, both curvatures are of $\OO(1/N_c)$, while the pion response enters
only at $\OO(1/N_c^2)$. Their sum is free of a Lorentz-invariant relative
vacuum term and can be directly compared with the hard and soft perturbative QCD
corrections computed in Ref.~\cite{Ferreira:2026qqu}. We keep the large-$N_c$ counting
separate from its ideal-gas realization, since planar interactions still
contribute at leading order.

The paper is organized as follows. In Section~\ref{sec:model} we define the
thermodynamic setup for the two phases. In
Section~\ref{sec:response} we present the response functions and the resulting
curvatures of the transition lines. Section~\ref{sec:results} contains
our results and their dependence on the renormalization scale and on the quark masses. In
Section~\ref{sec:chiral} we briefly compare our findings with those of chiral models.
The large-$N_c$ construction is presented in Sec.~\ref{sec:nc}, which
closes with a discussion of possible lattice tests.
Section~\ref{sec:conclusions} presents our conclusions.

\section{Thermodynamic setup}
\label{sec:model}
We work at zero baryon chemical potential and finite temperature $T$. For convenience, we write the electric and magnetic fields as 
$\cE=eE$ and $\cB=eB$, respectively. Here $e>0$ is the fundamental electric charge. 

For weak fields, the thermal free energy density of the matter in either phase can be expanded as
\begin{equation}
 f(T,E,B)=f_0(T)-\frac12\xi(T)\cE^2-\frac12\chi(T)\cB^2
 +\OO(F^4) \, ,
 \label{eq:potential}
\end{equation}
where the classical Maxwell term $(B^2-E^2)/2$ has been excluded and $F^4$ stands for terms $\sim (E^4, B^4, E^2 B^2)$. The dimensionless coefficients $\xi$ and $\chi$ refer to the static, spatially averaged, renormalized
response functions of Refs.~\cite{Endrodi:2022wym,Endrodi:2023wwf,Endrodi:2026kmb}, in which the static limit is taken
before the long-wavelength limit. These response functions must be computed from the photon polarization tensor in the presence of the non-dynamical external electric and magnetic fields~\cite{Endrodi:2022wym}. Notice that parity forbids an
$\bm E\cdot\bm B$ term in the CP-even medium at this order. In what follows we use $p=-f$ for the pressure that enters the
coexistence condition, without identifying it with every component of the mechanical stress tensor in a background field.

Now we consider two phases, a confined and a deconfined phase. In the spirit of Ref.~\cite{Fraga:2012fs}, we describe both phases by ideal gases. For each massive degree of freedom one has
\begin{equation}
 p_{B,F}(T,m)=\frac{1}{6\pi^2}\int_0^\infty\!\dd p\,
 \frac{p^4}{\omega_p}n_{B,F}(\omega_p) \, ,
 \label{eq:gas}
\end{equation}
where 
$\omega_p=\sqrt{p^2+m^2}$ 
and
$n_{B,F}(\omega)=[\exp(\omega/T)\mp1]^{-1}$ for bosons ($B$) and  fermions ($F$), respectively. The zero-field pressures of the two phases are then given by a free gas of massive pions (hadronic phase, $H$) and the MIT bag model for massive quarks plus free gluons~\cite{Chodos:1974je,Fraga:2012fs} (quark phase, $Q$):
\begin{align}
 p_{H,0}(T)&=3p_B(T,m_\pi) \, ,\nonumber\\
 p_{Q,0}(T)&=\frac{\pi^2}{45}(N_c^2-1)T^4
       +4N_c\sum_f p_F(T,m_f)-\Bbag \, ,
 \label{eq:phases}
\end{align}
where $N_c$ is the number of colors, $m_\pi$ is the pion mass,  $m_f$ is the $f$-flavor quark mass, and $\Bbag$ is the bag constant. 

The pion pressure includes the three real states, whereas its
electromagnetic response comes from a single complex charged scalar, so that
the charged pair is not counted twice. Each quark response includes particle
and antiparticle, with one Dirac spinor per color. Gluons have no direct electromagnetic response in this ideal gas setup. Here, all masses are taken to be independent of temperature.

In the calculation that follows, we choose $N_c=3$ and
\begin{gather}
 m_u=m_d=5~\mathrm{MeV},\qquad m_\pi=135~\mathrm{MeV},\nonumber\\
 Q_u=\frac23,\quad Q_d=-\frac13,\quad T_0\equiv\Tc(0)=162~\mathrm{MeV},
 \label{eq:parameters}
\end{gather}
where $Q_f$ is the quark charge in units of $e$. We adopt $T_0=162$~MeV inspired by the behavior of the Polyakov loop described in Ref. \cite{Endrodi:2023wwf}. The light quark masses enter as inputs, and the bag constant is fixed solely by the zero-field matching 
condition $p_{Q,0}(T_0)=p_{H,0}(T_0)$, which gives
\begin{equation}
 \Bbag^{1/4}\approx 226~\mathrm{MeV}
 \, ,\quad
 \Delta s(T_0)\approx 0.064~\mathrm{GeV}^3 
 \, ,
 \label{eq:bagvalue}
\end{equation}
where $\Delta X=X_Q-X_H$ and $s=\partial p/\partial T$ is the entropy. This value for the bag constant is about $10\%$ higher than the one in Ref. \cite{Fraga:2012fs}.

As discussed at length in Refs.~\cite{Fraga:2012fs,Endrodi:2013cs}, the treatment of vacuum contributions deserves special attention when different effective descriptions are used for the two phases. At one loop, each charged particle also contributes to the free energy a zero-temperature term quadratic in the fields, which is Lorentz invariant and ultraviolet divergent, and must be renormalized together with the electric charge. Expanding the constant-field determinants of Schwinger~\cite{Schwinger51} to second order in the fields, and renormalizing at the scale $\mu_0$, one finds for a Dirac fermion of charge $Q_f$ and for a complex scalar of
unit charge
\begin{align}
 f^{\rm vac}_F&=\frac{Q_f^2}{24\pi^2}\ln\frac{\mu_0^2}{m_f^2}\,(\cB^2-\cE^2),
 \nonumber\\
 f^{\rm vac}_S&=\frac{1}{96\pi^2}\ln\frac{\mu_0^2}{m^2}\,(\cB^2-\cE^2).
 \label{eq:vacuum}
\end{align}
The mass-independent parts of these terms, proportional to $\ln\mu_0^2$, are
absorbed by the renormalization of the charge. Since the external field and the electric charge are the same on both sides of the transition, the scale
$\mu_0$ must be the same in both phases. The logarithmic coefficients, on the other hand, differ, since the charged particles differ, so that the vacuum
terms do not cancel in the pressure balance. This is the weak-field counterpart of the subtleties of renormalization discussed in Ref.~\cite{Fraga:2012fs}. Notice that the subtraction of all mass-independent,
field-dependent terms adopted there for very strong fields cannot be carried over to the weak-field limit, where it would leave terms of order $B^2\ln|B|$ and no finite curvature: a fixed scale is required.

We fix $\mu_0$ by requiring that $e$ be the charge measured in the confined vacuum, in which the external fields are defined. For the pion gas this means $\mu_0=m_\pi$, so that its vacuum term vanishes. This mirrors the convention of lattice simulations, in which the susceptibilities are renormalized so that they vanish at zero temperature, consistently with unit
permittivity and permeability of the vacuum~\cite{Endrodi:2022wym}. The analogy is not complete, however: on the lattice there is a
  single vacuum, and no relative term arises. Here, $\mu_0$ is the common scale at
  which quarks and pions are renormalized and the relative vacuum term is evaluated,
  i.e.\ the scale at which the two descriptions are matched. In principle, it could be
  constrained by the hadronic vacuum polarization itself; its variation is discussed
  in Sec.~IV.
  
The deconfined phase
then acquires the relative vacuum term
\begin{equation}
 \begin{split}
 \delta f_Q&=\frac{c}{2}(\cB^2-\cE^2) \, ,\\
 c&=\frac{N_c}{12\pi^2}\sum_fQ_f^2\ln\frac{\mu_0^2}{m_f^2}
 -\frac{1}{48\pi^2}\ln\frac{\mu_0^2}{m_\pi^2} \, ,
 \end{split}
 \label{eq:matching}
\end{equation}
where the last term vanishes for $\mu_0=m_\pi$, but has been kept for later use. For the parameters adopted here, one finds $c= 0.0928$. Although $c$ contains
  $\ln(\mu_0^2/m_f^2)$ and thus looks sensitive to the current quark masses, this
  dependence cancels against the thermal logarithms of the quark response for
  $m_f\ll T$ (cf. Eq.~(\ref{eq:lightlimit}) below). Thus, the susceptibilities of the deconfined phase depend on $\ln(T/\mu_0)$, not on $\ln(\mu_0/m_f)$. Moreover, this relative vacuum term
  does not change the bag constant, since it vanishes at zero field, and {\it it is fixed by the particle content of the two phases, rather than being an adjustable parameter}. Its dependence on the choice of $\mu_0$ is discussed in Sec.~\ref{sec:results}. Notice also that, for quarks with mass $m_f=5$~MeV, the weak-field expansion of the vacuum determinant alone would require $|Q_f eF|\ll m_f^2$. This condition is too
  restrictive for the sum of vacuum and thermal contributions: the logarithms of $m_f$ cancel between them [Eq.~(\ref{eq:lightlimit})], so that for $m_f\ll T$ we expect the combined quark
  response to be controlled by $|Q_f eF|/T^2$, and the pion response by $eF/m_\pi^2$.
  At $T_0$ both scales are of order $0.02$--$0.03~\mathrm{GeV}^2$. The curvatures computed below are derivatives at zero field, and should be compared with lattice results obtained in, or extrapolated to, this regime.

\section{Response functions and transition curvatures}
\label{sec:response}

The electric and magnetic thermal susceptibilities correspond to second derivatives of the free energy density $f(T,E,B)$ with respect to the electromagnetic fields, and those can be expressed in terms of the second derivatives of the diagonal elements of the photon polarization tensor with respect to the momentum $k$ flowing through the loop ~\cite{Endrodi:2022wym,Endrodi:2026kmb}. For the electric susceptibility, one needs the second derivative of $\Pi_{00}^{T\neq 0}(k_0,k)$ and, for its magnetic counterpart, the second derivative of $\Pi_{ii}^{T\neq 0}(k_0,k)$. In both cases, the $T=0$ contribution has already been subtracted from the polarization tensor. Following Refs.~\cite{Endrodi:2022wym,Endrodi:2026kmb}, after taking the limits $k_0\to 0$ and $k\to 0$ {\it in this order}, one finds
\begin{align}
 \xi_S(T,m)&=\frac{1}{12\pi^2}\int_m^\infty\!
 \frac{\dd\omega}{\sqrt{\omega^2-m^2}}
 [n_B(\omega)+\omega n_B'(\omega)] \, ,\label{eq:scalarE}\\
 \chi_S(T,m)&=-\frac{1}{12\pi^2}\int_m^\infty\!
 \frac{\dd\omega}{\sqrt{\omega^2-m^2}}n_B(\omega) \, ,
 \label{eq:scalarB}
\end{align}
where the prime on $n$ denotes a derivative with respect to the energy. The corresponding expressions for a Dirac fermion read~\cite{Endrodi:2022wym}
\begin{align}
 \xi_F(T,m)&=-\frac{1}{6\pi^2}\int_m^\infty\!
 \frac{\dd\omega}{\sqrt{\omega^2-m^2}}
 [2n_F(\omega)-\omega n_F'(\omega)] \, ,\label{eq:fermionE}\\
 \chi_F(T,m)&=\frac{1}{3\pi^2}\int_m^\infty\!
 \frac{\dd\omega}{\sqrt{\omega^2-m^2}}n_F(\omega) \, .
 \label{eq:fermionB}
\end{align}
Adding the vacuum terms of Eq.~\eqref{eq:vacuum}, the total response functions of the two phases are given by
\begin{equation}
 \begin{split}
 (\xi_H,\chi_H)&=(\xi_S+\ell_S,\;\chi_S-\ell_S) \, ,\\
 (\xi_Q,\chi_Q)&=N_c\sum_fQ_f^2
 (\xi_F+\ell_f,\;\chi_F-\ell_f) \, ,
 \end{split}
 \label{eq:quarkresp}
\end{equation}
where the thermal responses are evaluated at $m=m_\pi$ and $m=m_f$, respectively, and 
\begin{eqnarray}
\ell_f&=& \frac{1}{12\pi^2}\log \frac{\mu_0^2}{m_f^2} \, ,
\label{eq:ellf-N}
\\
\ell_S&=& \frac{1}{48\pi^2}\log \frac{\mu_0^2}{m_\pi^2} \, ,
\end{eqnarray}
so that $\ell_S=0$ for $\mu_0=m_\pi$.
Equations~\eqref{eq:scalarE}--\eqref{eq:fermionB} are known single-phase results; what we need here is their effect on the pressure crossing. Notice, however, that the electric scalar integral contains a subtraction that is implicit in the long-wavelength limit.

The signs of the individual response functions follow immediately. Both phases have a negative electric susceptibility: e.g., for the scalar, setting $x=\omega/T$ one
finds $n_B+\omega n_B'=n_B[1-x(1+n_B)]<0$, since $x>1-e^{-x}$. The magnetic response, on the other hand, is diamagnetic for the scalar and paramagnetic for the fermion. Their sums satisfy
\begin{align}
 \xi_S+\chi_S&=\frac{1}{12\pi^2}\int_m^\infty\!
 \frac{\dd\omega\,\omega n_B'}{\sqrt{\omega^2-m^2}}<0 \, ,\nonumber\\
 \xi_F+\chi_F&=\frac{1}{6\pi^2}\int_m^\infty\!
 \frac{\dd\omega\,\omega n_F'}{\sqrt{\omega^2-m^2}}<0 \, .
 \label{eq:sums}
\end{align}

Notice that for $m/T\ll1$, one has $\xi_F+\chi_F\to-1/(12\pi^2)$, even
though the separate fermion coefficients contain logarithms of $T/m$. The
scalar sum, instead, has an infrared enhancement of order $T/m$, which is why we keep the pion mass finite also when discussing the limit of large number of colors in the sequel.

The vacuum terms modify this picture in an important fashion. Since they are Lorentz invariant, they cancel in the sums of Eq.~\eqref{eq:sums}, but their logarithms of the mass cancel those of the thermal fermion response. For $m\ll T$, one finds
\begin{equation}
 \chi_F-\ell_f\;\longrightarrow\;\frac{1}{12\pi^2}
 \left[\ln\frac{\pi^2T^2}{\mu_0^2}-2\gamma_E\right] 
 \label{eq:lightlimit}
\end{equation}
and
\begin{equation}
\xi_F+\ell_f=-\frac{1}{12\pi^2}-(\chi_F-\ell_f) \, , 
\end{equation}
which is the leading-order perturbative result~\cite{Ferreira:2026qqu}. For $m_f=5$~MeV, Eq.~\eqref{eq:lightlimit} is accurate to about $10^{-4}$. Thus, the quark phase remains paramagnetic as long as $T>e^{\gamma_E}\mu_0/\pi\simeq0.57\,\mu_0$, but its susceptibilities are much smaller than the thermal ones alone.

Expanding the coexistence condition $p_Q(\Tc,E,B)=p_H(\Tc,E,B)$ about $T_0$,
we obtain
\begin{align}
 \Tc(E,B)&=T_0+\frac12\kappa_E\cE^2
                  +\frac12\kappa_B\cB^2+\OO(F^4) \, ,\label{eq:tc}\\
 \kappa_E&=-\frac{\Delta\xi(T_0)}{\Delta s(T_0)} \, ,\qquad
 \kappa_B=-\frac{\Delta\chi(T_0)}{\Delta s(T_0)} \, ,
 \label{eq:curvatures}
\end{align}
where the factor $1/2$ fixes our convention for the curvatures, both of
which have units of $\mathrm{GeV}^{-3}$. Equation~\eqref{eq:curvatures} is
simply a coexistence identity, valid whenever the zero-field crossing has a
nonzero entropy jump; the model determines its numerator and denominator
separately.

The physical picture is then quite simple. For light quarks one has
$\xi_Q<\xi_H<0$, so that the electric field raises the free energy of the
quark phase more strongly, favoring the pion gas at the original crossing,
and a higher temperature is required for deconfinement. On the other hand,
$\chi_Q>0>\chi_H$, so that the magnetic field lowers the free energy of the quark
phase and raises that of the pion gas, and the transition moves to lower
temperatures. Notice that the electric sign requires a comparison of
magnitudes, since the fact that both susceptibilities are negative is not
sufficient by itself.

\section{Results}
\label{sec:results}

\begin{figure*}[t]
\includegraphics[width=\textwidth]{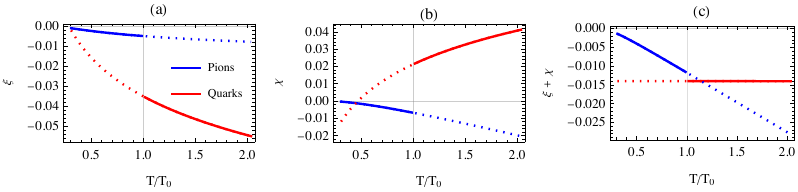}
\caption{Electric and magnetic susceptibilities, and their sum, as functions of the temperature, including the vacuum terms with $\mu_0=m_\pi$. Solid lines follow the phase with
higher zero-field pressure, while dotted lines continue the separate gas branches. The discontinuity reflects the bag construction, and no crossover
interpolation is introduced. The sum $\xi+\chi$ does not depend on the vacuum terms.}
\label{fig:response}
\end{figure*}

Before discussing the transition itself, it is instructive to look at the behavior of the response functions of the two phases, as shown in Fig.~\ref{fig:response}. The figure illustrates how the change in the
charged degrees of freedom across the transition produces the two different
response functions. At low temperatures the pion expressions reproduce the analytic result of Ref.~\cite{Endrodi:2026kmb}, where the pion contribution was also compared with lattice data. Choosing $\mu_0=m_\pi$, they carry no vacuum term. In the quark phase, instead, the vacuum term shifts $\xi_Q$ and $\chi_Q$. As shown in Eq.~\eqref{eq:lightlimit}, the remaining temperature
dependence is only logarithmic. The change from diamagnetism at low temperatures to paramagnetism at high temperatures, as shown in the behavior of $\chi$ in the figure, has also been observed on the lattice~\cite{Bali:2020bcn}, although our simple framework obviously cannot reproduce the smooth transition region. 

At the matching point for the pressures of the two phases, we find
\begin{equation}
 \kappa_E=0.473~\mathrm{GeV}^{-3},\qquad
 \kappa_B=-0.437~\mathrm{GeV}^{-3},
 \label{eq:referencek}
\end{equation}
with the corresponding response functions for each phase listed in Table~\ref{tab:response}.
Notice that {\it no electromagnetic coefficient has been adjusted: besides
the masses, the only input is the renormalization condition that fixes $\mu_0$}. 

The table also shows that the relative vacuum term is of approximately the same size as the thermal response functions, and of opposite sign. So, it reduces both curvatures by a factor of about four\footnote{The values for the response functions without including the vacuum contribution would be $\kappa_E=1.923$~GeV$^{-3}$ and $\kappa_B=-1.887$~GeV$^{-3}$.}. The resulting electric curvature
is consistent with the lattice value $\kappa_E=0.37(9)~\mathrm{GeV}^{-3}$
extracted from the Polyakov loop in Ref.~\cite{Endrodi:2023wwf}. For the magnetic
curvature, a weak-field fit to the lattice data of Ref.~\cite{Bali:2011qj} gives
$T_c(eB)/T_c(0)=1-v_B(eB)^2$, with $v_B=0.55(5)~\mathrm{GeV}^{-4}$ at the physical pion mass~\cite{DElia:2018xwo}. Using our value for $T_0$, this corresponds to
$\kappa_B=-2T_0v_B=-0.178(16)~\mathrm{GeV}^{-3}$. Alternatively, the
continuum transition temperatures of Ref.~\cite{Bali:2011qj} have been
interpolated in Ref.~\cite{Endrodi:2015oba} as
$T_c(eB)=T_c(0)[1+a_1(eB)^2]/[1+a_2(eB)^2]$. Expanding for small fields,
$T_c(eB)/T_c(0)\simeq1+(a_1-a_2)(eB)^2$, so that $\kappa_B=2T_0(a_1-a_2)$. With our $T_0$ and treating the errors of $a_1$ and $a_2$ as independent,
this gives $\kappa_B=-0.091(9)~\mathrm{GeV}^{-3}$ for the light-quark condensate and $-0.162(5)~\mathrm{GeV}^{-3}$ for the strange quark number
susceptibility. Notice, however, that this interpolation was designed to describe
fields up to $eB=3.25~\mathrm{GeV}^2$, rather than the weak-field limit.
Our magnetic curvature, therefore, has the right sign but is between 2.5
and 5 times larger in magnitude than these lattice estimates. Nevertheless,
these quantitative comparisons should not be overemphasized: the lattice
results describe a crossover
with physical flavor content, while our framework has two
flavors and a first-order phase transition. Moreover, as we discuss below,
the curvatures depend on the choice of the renormalization scale.

\begin{table}[t]
\caption{Susceptibilities at $T_0=162$ MeV, with
$\mu_0=m_\pi$. The first two rows give the thermal responses, the third the
vacuum contribution to $\xi$, Eq.~\eqref{eq:vacuum} (that to $\chi$ has the
opposite sign), and the last three rows the total responses. All entries
are dimensionless, and the last column gives the difference between the
deconfined and the hadronic responses.}
\label{tab:response}
\begin{ruledtabular}
\begin{tabular}{lrrr}
 & Pions & Quarks & Difference\\
$\xi^{\rm th}$ & $-0.00493$ & $-0.128$ & $-0.123$\\
$\chi^{\rm th}$ & $-0.00685$ & $0.114$ & $0.121$\\
$\xi^{\rm vac}=-\chi^{\rm vac}$ & $0$ & $0.0928$ & $0.0928$\\
$\xi$ & $-0.00493$ & $-0.0352$ & $-0.0302$\\
$\chi$ & $-0.00685$ & $0.0211$ & $0.0280$\\
$\xi+\chi$ & $-0.0118$ & $-0.0141$ & $-0.00229$\\
\end{tabular}
\end{ruledtabular}
\end{table}

\begin{figure}[t]
\includegraphics[width=\columnwidth]{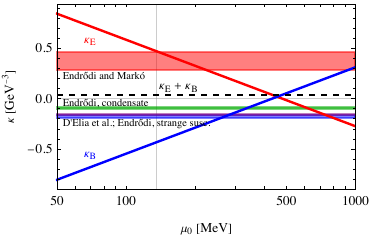}
\caption{Electric and magnetic curvatures as functions of the renormalization scale
$\mu_0$ at which the relative vacuum term, Eq.~\eqref{eq:matching}, is
evaluated. The vertical line marks our choice $\mu_0=m_\pi$. Bands
indicate the lattice values $\kappa_E=0.37(9)~\mathrm{GeV}^{-3}$
(red)~\cite{Endrodi:2023wwf} and, for the magnetic curvature, $-0.178(16)$ (blue), from
the fit of Ref.~\cite{DElia:2018xwo} to the data of Ref.~\cite{Bali:2011qj}, and
$-0.091(9)$ (green) and $-0.162(5)~\mathrm{GeV}^{-3}$ (purple), from the
small-field expansion of the interpolation of Ref.~\cite{Endrodi:2015oba} for the
light-quark condensate and the strange quark number susceptibility,
respectively. The sum $\kappa_E+\kappa_B$ (dashed) does not depend on
$\mu_0$.}
\label{fig:scale}
\end{figure}

The choice $\mu_0=m_\pi$ deserves some comment. Notice that the logarithmic
coefficients of the two phases differ. Thus, changing $\mu_0$
while keeping the finite part of the relative vacuum term equal to zero at
the new scale defines a different matching between the two descriptions,
rather than a renormalization-group trajectory, and the particle spectrum
alone does not fix it. Our choice is singled out by the requirement that
the charge be that of the confined vacuum, in which the hadronic
susceptibilities vanish at zero temperature. To illustrate the sensitivity
to this choice, Fig.~\ref{fig:scale} shows the electric and magnetic curvatures as
functions of $\mu_0$. For $m_\pi\le\mu_0\le T_0$ the electric curvature
varies between $0.473$ and $0.405~\mathrm{GeV}^{-3}$. The signs survive up
to $\mu_0=434$ MeV for the magnetic curvature and $478$ MeV for the
electric one; only for scales well above the transition temperature, where
the vacuum term overwhelms the thermal paramagnetism of the quarks, does
the magnetic curvature change sign. The lattice magnetic curvatures would
be reproduced for $\mu_0$ between 270 and 340 MeV, where, however,
$\kappa_E$ lies between $0.13$ and $0.21~\mathrm{GeV}^{-3}$, below the
lattice electric value: both cannot be matched at the same scale, as the
sum of the curvatures shows.

Interestingly, the two curvatures almost cancel:
\begin{equation}
 \kappa_E+\kappa_B=-\frac{\Delta(\xi+\chi)}{\Delta s}
 =0.0358~\mathrm{GeV}^{-3} \, .
 \label{eq:curvsum}
\end{equation}
Since the vacuum terms are Lorentz invariant, this sum does not depend on
them, and therefore neither on $\mu_0$ nor on a common electromagnetic
counterterm. It provides a useful observable to compare descriptions with
different vacuum conventions, provided their thermal prescriptions agree.
It should not be regarded, however, as a universal order parameter: it need
not vanish in either phase, its jump has no fixed sign at finite $N_c$, and
a temperature-dependent matching potential with independent electric and
magnetic coefficients would not cancel in it. Its small value reflects the
small jump in $\xi+\chi$, rather than any analytic continuation relating
the thermal electric and magnetic problems. 
The lattice values quoted
above give $\kappa_E+\kappa_B$ between $0.19(9)$ and
$0.28(9)~\mathrm{GeV}^{-3}$, although they were obtained with different observables and discretizations. Unlike the
individual curvatures, this difference cannot be attributed to the vacuum matching, and points instead to the limitations of the ideal-gas thermal responses near the transition.

One could wonder how general are the signs of the individual curvatures. For the magnetic
one, an argument can be given within this class of models. The 16 massless gluon degrees of freedom alone already carry more entropy at $T_0$ than the three massive pion states retained in the hadronic phase, and quarks only
  add to it, so that $\Delta s>0$. For light quarks,
Eq.~\eqref{eq:lightlimit} shows that $\chi_Q>0$ as long as
$T_0>0.57\,\mu_0$, while the pion gas is diamagnetic. Therefore
$\Delta\chi>0$ and $\kappa_B<0$. For the electric curvature, notice that the sum $\xi_Q+\chi_Q$ is fixed by the thermal response alone, and tends to $-N_c\sum_fQ_f^2/(12\pi^2)$ for light quarks. A paramagnetic quark phase then implies $\xi_Q<-N_c\sum_fQ_f^2/(12\pi^2)$, which for two flavors is about three times larger in magnitude than the pion response at $T_0$, so
that $\Delta\xi<0$ and $\kappa_E>0$.

\begin{figure}[t]
\includegraphics[width=\columnwidth]{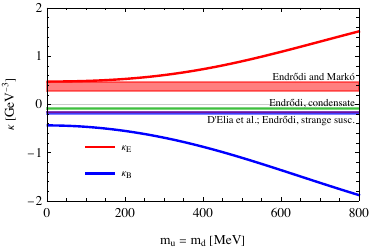}
\caption{Transition curvatures as functions of the degenerate quark mass
for two flavors, with $T_0=162$ MeV, $m_\pi=135$ MeV and $\mu_0=m_\pi$; only
the zero-field bag constant is redefined at each mass so that $T_c(0)=T_0=162$ MeV. The bands indicate, for
comparison, the lattice values of $\kappa_E$ (red)~\cite{Endrodi:2023wwf} and
$\kappa_B$ (blue, green and purple)~\cite{Bali:2011qj,DElia:2018xwo,Endrodi:2015oba}, as in
Fig.~\ref{fig:scale}. Both curves are zero-field
derivatives, so that no range of validity in the fields is implied.}
\label{fig:curvatures}
\end{figure}

The dependence on the quark mass is shown in Fig.~\ref{fig:curvatures},
where only the zero-field bag constant is redefined at each mass so that $T_c(0)=T_0=162$ MeV. Both curvatures
keep their signs over the whole range considered. Thanks to the
cancellation of logarithms in Eq.~\eqref{eq:lightlimit}, the light-mass
limit is smooth: between 5 and 2.5 MeV the curvatures change by about
$10^{-4}$ in relative terms. For heavier quarks the thermal responses
decrease, but the vacuum logarithm in Eq.~\eqref{eq:matching} changes sign
for $m_f>\mu_0$, so that both curvatures grow in magnitude, reaching
$\kappa_E=1.51$ and $\kappa_B=-1.88~\mathrm{GeV}^{-3}$ at 800 MeV. In this
regime the result is dominated by the vacuum matching, rather than by the
thermal physics.

\section{Comparison with chiral models}
\label{sec:chiral}

The simultaneous signs found above arise from comparing the field-dependent free energies of two phases with different charged fields, a separation that the bag model imposes directly. Of course, the model neither solves a chiral gap equation nor includes the response of the gauge fields to the
sea quarks. Its success indicates that this simple thermodynamic mechanism is sufficient to account for the qualitative trend, but it does not show that the omitted dynamics is irrelevant. In this sense, our findings are in line with the suggestion of Refs.~\cite{Fraga:2012fs,Fraga:2012ev} that confinement might play a central role in guiding the behavior of the critical temperature in external fields.

For magnetic fields, lattice studies have identified a sea-quark
contribution that modifies the distribution of gauge fields and can overcome the valence enhancement of the chiral condensate near the crossover~\cite{Bruckmann:2013oba}. Several chiral models incorporate the missing field dependence through fitted couplings. For instance, the scalar coupling was
fitted to the magnetic pseudocritical curve in Ref.~\cite{Ferreira:2014kpa},
whereas in Ref.~\cite{Farias:2014eca} an Ansatz for its magnetic and thermal
dependence was constrained using condensate data. While the first procedure takes the transition curve as an input, the second supplies a related thermodynamic constraint, and a prescription calibrated to zero-temperature baryon information~\cite{Endrodi:2019whh} has an even greater independence from the thermal transition. These different uses of external
information should be kept in mind when assessing the predictive content of each approach.

Notice that a fitted field dependence can modify the curvature already at
leading order in weak fields. If $z=(eF)^2$, $G(z)=G_0+g_2z+\cdots$, and
$\sigma$ is a mean field, one has, at fixed $\sigma$, that the thermodynamic potential $\Omega$ is such that
\begin{equation}
 \left.\frac{\partial\Omega}{\partial z}\right|_{z=0}
 =\Omega_2(T,\sigma;G_0)
   +g_2\frac{\partial\Omega_0(T,\sigma;G_0)}{\partial G} \, .
 \label{eq:coupling}
\end{equation}
At a stationary solution, the implicit first derivative of $\sigma$ does not remove this explicit coupling term. Fitting it to a transition shift therefore removes that shift as an independent test, although other
predictions may remain.

We should remark, however, that a decreasing effective four-fermion coupling is not intrinsically unphysical: unlike the QCD gauge coupling, it is dimensionful
and depends on the cutoff and on the operators retained. More useful tests are provided by its analytic structure at weak fields and by its range of validity. In the parametrization of Ref.~\cite{Farias:2014eca}, for instance,
\begin{equation}
 \frac{G(B,T)}{G_0}=
 \frac{1-\gamma |eB|T/\Lambda^3}
 {1+\alpha\ln[1+\beta |eB|/\Lambda^2]} \, ,
 \label{eq:farias}
\end{equation}
with positive coefficients, and the term linear in $|B|$ produces a cusp at zero field. Such a behavior is incompatible with the smooth, even Taylor expansion in weak fields assumed here away from a critical singularity. Of course, a parametrization designed for large fields can still be useful within its fitted range, but its motivation from asymptotic freedom does not determine the weak-field
response. Moreover, nonlocal chiral models with fixed parameters can exhibit inverse magnetic catalysis~\cite{Pagura:2016pwr}, so that the magnetic problem does not provide a general no-go theorem for chiral descriptions
without a fitted field-dependent coupling, even though one has now to deal with nonlocal models.

For electric fields, the conventional NJL/PNJL and quark--meson implementations of Refs.~\cite{Tavares:2019mvq,Tavares:2023ybt} (cf. also \cite{Cao:2015dya}) find an initial decrease of the pseudocritical temperature, opposite to the lattice trend. These studies, however, employ proper-time electric constructions and chiral or Polyakov-loop criteria, which do not correspond to the coexistence problem
and the static prescription considered here. Our electric susceptibilities employ the static weak-field prescription underlying the lattice analysis. This facilitates comparison of the electromagnetic response, although our pressure-crossing transition differs from the lattice Polyakov-loop crossover criterion. Several existing chiral-model calculations use a different electric-field prescription; their disagreement with the lattice trend therefore does not establish a general failure of fixed-coupling chiral models. As emphasized in Ref.~\cite{Endrodi:2023wwf}, the behavior under an alternative electric prescription remains an open issue.

\section{Large-$N_c$ description}
\label{sec:nc}

Let us now turn to the analysis of the deconfinement transition in weak electric and magnetic 
backgrounds in the large $N_c$ limit of QCD. As shown in Ref.~\cite{Fraga:2012ev} for the magnetic case, the large-$N_c$
limit provides a well-defined setup in which the pressure balance can be
organized by essentially counting powers of $1/N_c$. In this regime, the qualitative behavior driven by deconfinement dynamics associates paramagnetism of quarks with the phenomenon of inverse magnetic catalysis, characterized by a negative curvature $\kappa_B$. Here, we extend the large-$N_c$
analysis to the static electric prescription used throughout this work, showing that both curvatures are of $O(1/N_c)$.
We first derive the general large-$N_c$ results,
then identify the additional assumptions under which the electric
curvature is positive. Finally, we retain the expansion through order
$1/N_c^2$ in both the zero-field transition temperature and the
curvatures, and evaluate the resulting expressions in the specific case of our framework.

\subsection{Color counting and the physical content of the pressure balance}
\label{subsec:lnc-counting}

Let us consider the 't~Hooft limit at fixed $N_f$, fixed
$\lambda=g_s^2N_c$, and vanishing baryon chemical
potential~\cite{tHooft:1973alw,Witten:1979kh}. Quark and meson masses,
temperature, and the external fields $\mathcal E=eE$ and
$\mathcal H=eB$ remain of order $N_c^0$. Unless stated otherwise, the charges $Q_f$ are fixed;
their alternative color dependence will be specified below. The
expansion is taken at nonzero masses and fixed mass-to-temperature
ratios.

We assume smooth phase branches near a first-order crossing with a
nonzero entropy jump, as appropriate to the large-$N_c$
pure-glue transition~\cite{Lucini:2012wq}. We adopt the notation:
$X_E=\xi$ and $X_B=\chi$ for $\mathcal F=\mathcal E,\mathcal H$,
respectively. The pressure difference between confined and deconfined phases, $\Delta p=p_Q-p_H$, in the presence of one nonzero electromagnetic field reads
\begin{align}
 \Delta p(T,\mathcal F)
 &=\Delta p_0(T)+\frac{\mathcal F^2}{2}\Delta X_F(T)
       +O(\mathcal F^4) \, ,  \label{eq:lnc-pressure}
 \end{align}
where the zero-field pressure difference is
 \begin{align}
 \Delta p_0(T)
 &=N_c^2 A(T)+N_c Q(T)+R(T)+O(N_c^{-1}) \, , 
 \label{eq:0fieldDp}
 \end{align}
 and the \emph{net} electromagnetic response is encoded by
 \begin{align}
 \Delta X_F(T)
 &=N_c r_F(T)+j_F(T)+O(N_c^{-1}) \, .
\end{align}
The functions $A,Q,R$ have dimensions of pressure, while $r_F,j_F$
are dimensionless. The pressure is the reduced thermodynamic
potential $p=-f$ already defined above, within the static equilibrium
prescription in the case of electric fields.

The leading term $N_c^2A$ contains the pure-glue pressure difference,
including planar interactions and the relative confinement vacuum
energy. The confined \emph{thermal} pressure is $O(1)$, but the
confined vacuum energy need not be. A common vacuum normalization
cancels from the pressure difference, while a relative confinement energy
remains. Thus $A(T)$ is not restricted to a Stefan--Boltzmann
pressure minus a constant.

The flavor coefficient $N_cQ$ contains both the deconfined thermal
flavor pressure, with planar gluon dressing, and quark-induced changes
in the relative vacuum or confinement energy\footnote{The symbol $Q$ here does
not denote an electric charge, and its net sign is not fixed by color
counting.}.  The $O(N_c^0)$ term $R$ includes confined thermal mesons and
glueballs, subleading gluon contributions, and diagrams with two quark
loops. Baryons have masses of order $N_c$ and are thermally
suppressed in the regime considered. 

Similarly, $N_cr_F$ is the leading \emph{net} electromagnetic response
difference. It includes charged quark loops, their planar gluon
dressing, and any leading relative vacuum response. Since gluons are
electromagnetically neutral, there is no purely gluonic
$O(N_c^2)$ susceptibility. Nevertheless, virtual quarks can change
confinement dynamics and the confined vacuum response at $O(N_c)$.
The coefficient $j_F$ contains subleading responses, including the confined thermal susceptibility with a minus sign. Contributions from
two quark loops terms are also allowed at
this order.

\subsection{Critical temperature and curvatures}
\label{subsec:lnc-monotonicity}

The leading, pure-glue deconfinement transition is determined by 
\begin{equation}
 A(T_g)=0 \, ,
\label{eq:lnc-glue-crossing}
\end{equation}
and is characterized by positive entropy generation, so that $a_1=A'(T_g)>0$. 
Let $T_N\equiv T_c(0;N_c)$ denote the zero-field crossing at finite number of colors.
 If
$T_N=T_g+t_1/N_c+\cdots$, the order-$N_c$ part of
$\Delta p_0(T_N)=0$ gives $a_1t_1+Q(T_g)=0$, or
\begin{equation}
 T_N=T_g-\frac{Q(T_g)}{N_ca_1}+O(N_c^{-2}).
 \label{eq:lnc-leading-crossing}
\end{equation}
Flavors therefore lower the zero-field transition temperature if
$Q(T_g)>0$. This seems reasonable, given the fact that lattice results support a reduction of the thermal transition
temperature upon including dynamical quarks, both at $N_c=3$
\cite{KarschLaermannPeikert2001} and in exploratory simulations at $N_c=4,5$ \cite{DeGrand2021}. Still, we treat the sign of the complete flavor contribution as an additional input, even though the order of the shift is fixed.

At fixed $N_c$, the electric/magnetic curvature $\kappa_F$ is directly obtained from the critical temperature
\begin{equation}
 T_c(\mathcal F;N_c)
 =T_N+\frac12\kappa_F(N_c)\mathcal F^2+O(\mathcal F^4).
 \label{eq:lnc-curvature-definition}
\end{equation}
Expanding the coexistence condition ($\Delta p (T_c,\mathcal{F})=0$) around $T_N$ yields
\begin{equation}
 0=\Delta s(T_N)\,[T_c-T_N]
       +\frac{\mathcal F^2}{2}\Delta X_F(T_N)
       +O(\mathcal F^4),
 \label{eq:lnc-balance-expansion}
\end{equation}
where $\Delta s=\partial_T\Delta p_0$. The temperature derivative of
the susceptibility first contributes at quartic field order, since it
multiplies both $T_c-T_N$ and $\mathcal F^2$. Hence
\begin{equation}
 \kappa_F=-\frac{\Delta X_F(T_N)}{\Delta s(T_N)}
          =-\frac{r_F(T_g)}{N_ca_1}+O(N_c^{-2}) \, .
 \label{eq:lnc-coexistence-identity}
\end{equation}
The curvatures have units of inverse energy cubed. Their generic
$1/N_c$ suppression follows from dividing a response of order $N_c$
by an entropy jump of order $N_c^2$.

The physical sign argument is immediate once the \emph{matched}
susceptibility $\Delta X_F(T_N)$ difference is known. For $\Delta s>0$, a negative
$\Delta\xi$ raises $f_Q-f_H$ at the original crossing, favoring the
confined phase. A higher temperature is then required to restore
pressure equality, so $\kappa_E>0$. Positive $\Delta\chi$ instead
favors the deconfined phase and gives $\kappa_B<0$. The sign of either
response in one phase alone is insufficient at finite $N_c$.

This reasoning also gives a local monotonicity statement without specifying a gluonic equation of state. Setting $z=\mathcal E^2$ and
differentiating $\Delta p(T_c(z),z)=0$ yields
\begin{equation}
 \frac{dT_c}{dz}
 =-\left.\frac{\partial_z\Delta p}
                  {\partial_T\Delta p}\right|_{T=T_c(z)} \, .
 \label{eq:lnc-monotonic-slope}
\end{equation}
Whenever $\partial_T\Delta p>0$ and $\partial_z\Delta p<0$ along a
smooth coexistence branch, $T_c$ increases with $z$. At zero field,
$\partial_z\Delta p=\Delta\xi/2$ and
$\kappa_E=2(dT_c/dz)_0$. Thus $r_E(T_g)<0$ implies positive leading
electric curvature for any $A(T)$ with a positive entropy jump at the transition. 
The sign of the electric response therefore plays the
same indispensable role as the paramagnetic
response in the magnetic analysis.

The leading shifts of the transition temperature at large-$N_c$ in the presence of an electric field read
\begin{equation}
 T_c(\mathcal E)
 =T_g-\frac{Q(T_g)}{N_ca_1}
       -\frac{r_{E}(T_g)\mathcal E^2}{2N_ca_1}
       +O(N_c^{-2})+O(\mathcal E^4/N_c) \, .
 \label{eq:lnc-combined-shift}
\end{equation}
If the flavor contribution to the pressure balance evaluated at the pure-glue transition temperature is positive, i.e. $N_c Q(T_g)>0$, and $r_{E,g}<0$, the electric field partly reverses the zero-field
flavor decrease, with $T_N<T_c(\mathcal E)<T_g$ for sufficiently small nonzero field and sufficiently large $N_c$. This does not guarantee however a saturation of the transition temperature below the pure-glue value, since Eq. (\ref{eq:lnc-combined-shift}) is only valid at small fields.

To obtain the two curvatures at next-to-next-to-leading order, we write
\begin{equation}
 \Tc=T_g+\frac{t_1}{N_c}+\frac{t_2}{N_c^2}+\cdots,\qquad
 \kappa_F=\frac{k_{1F}}{N_c}+\frac{k_{2F}}{N_c^2}+\cdots.
 \label{eq:series}
\end{equation}
Evaluating all functions on the right-hand side at $T_g$, one finds
\begin{align}
 t_1&=-\frac{Q}{a_1},&
 t_2&=-\frac{a_2t_1^2/2+Q't_1+R}{a_1},\nonumber\\
 k_{1F}&=-\frac{r_F}{a_1},&
 k_{2F}&=-\frac{j_F+t_1r_F'}{a_1}
 +\frac{r_F(a_2t_1+Q')}{a_1^2}.
 \label{eq:nlo}
\end{align}
In the free model with fixed charges, $j_E=-\xi_H$ and $j_B=-\chi_H$.

\subsection{Vacuum response and the curvature signs}
\label{subsec:lnc-vacuum-signs}

The relative vacuum energy of the flavor sector may also play a role in this regime. Here, it is encoded in $Q$ together with the thermal flavor pressure, and similarly the field responses may be written as:
\begin{equation}
 r_E=\sum_f Q_f^2(\xi_f+\ell_f),\qquad
 r_B=\sum_f Q_f^2(\chi_f-\ell_f),
 \label{eq:ncr}
\end{equation}
where $\ell_f$ is the vacuum term of Eq.~\eqref{eq:ellf-N} per color. In
the interacting theory the flavor functions also include planar gluon
dressing, as well as possible field-dependent corrections to the bag
constant or to the string tension. Notice that the counting fixes the
$1/N_c$ suppression, but not the signs in Eq.~\eqref{eq:ncr}.

In our free realization, with $\mu_0=m_\pi$, Eq.~\eqref{eq:lightlimit}
shows that the leading electric curvature is positive and the magnetic one
is negative as long as $T_g>0.57\,\mu_0$, while the pion response, being of
$\OO(1)$, contributes to the curvature only at order $1/N_c^2$, with no
vacuum term at this scale. Notice that the relative vacuum term of the
quarks is not suppressed at large $N_c$: it scales as $N_c$, like their
thermal response, and its effect, therefore, does not disappear in this
limit. In full
QCD, however, even in  the large-$N_c$ limit, 
two-quark-loop terms and additional confined states can also
enter at this order, so that evaluating Eq.~\eqref{eq:nlo} with three pions
does not amount of course  to a full next-to-next-to-leading-order QCD calculation.

Further information can only be obtained by specifying more dynamical assumptions.
In the non-interacting case, the free integrals give $\xi_f<0$, $\chi_f>0$, and
\begin{equation}
 \xi_f+\chi_f=-\frac{T}{2}\frac{\partial\chi_f}{\partial T}<0
 \quad\text{at fixed masses}.
 \label{eq:lnc-thermal-identity}
\end{equation}
As temperature rises, thermally populated quarks strengthen the
magnetic response, while the magnitude of their negative electric response becomes still larger. Consequently, with $k_{1F}=-r_F(T_g)/a_1$,
\begin{equation}
 k_{1E}+k_{1B}
 =-\frac{\sum_fQ_f^2[\xi_f(T_g)+\chi_f(T_g)]}{a_1}>0.
 \label{eq:lnc-leading-sign-sum}
\end{equation}
In this sum an exact cancellation of the vacuum terms occurs as a direct consequence of cancellation previously pointed out for $\xi+\chi$ \cite{Endrodi:2022wym}. Negative magnetic curvature therefore
implies a larger positive electric curvature:
$k_{1B}<0$ requires
$k_{1E}>|k_{1B}|$.
Both leading curvatures cannot be negative in this realization.
For interacting QCD, the same implication requires
$r_E(T_g)+r_B(T_g)<0$, but its validity near deconfinement is an
open question that could in principle be further investigated on the lattice.

For light quarks, the adopted prescription yields $\kappa_E>0$
at leading order for $T_g>e^{\gamma_E-1/2}\mu_0/\pi\simeq0.344\mu_0$,
and $\kappa_B<0$ for $T_g>e^{\gamma_E}\mu_0/\pi\simeq0.567\mu_0$.
The latter condition suffices for both signs. These thresholds
depend on both the light-mass approximation and vacuum prescription.

\subsection{A fixed-bag realization}

In order to obtain quantitative results and comparison, we adopt now the specific bag-model we have discussed earlier and analyze its behavior as one varies the number of colors and in the large-$N_c$ expansions derived above.

To connect this counting with the $N_c=3$ model, we take the bag constant for varying number of colors as
\begin{align}
 \Bbag(N_c)&=(N_c^2-1)b_g-\alpha N_cN_f b_g+N_c b_m,
 \end{align}
so that the $O(N_c^2)$, $O(N_c)$ and $O(N_c^0)$ coefficients in the zero-field pressure difference (Eq. (\ref{eq:0fieldDp})) read, respectively ($a\equiv \frac{\pi^2}{45}$): 
\begin{align} 
A(T)&=aT^4-b_g,
\nonumber\\
 Q(T)&=4\sum_f p_F(T,m_f)+\alpha N_f b_g-b_m,\nonumber\\
 R(T)&=-aT^4+b_g-p_{H,0}(T).
 \label{eq:ncmodel}
\end{align}
The parameters $\alpha$ and $b_m$ are independent inputs, representing,
respectively, a zero-field reduction due to the flavors, analogous to the
reduction of the string tension in Ref.~\cite{Fraga:2012ev}, and a mass-dependent
contribution to the relative vacuum energy. For our baseline we set both to
zero and choose $b_g=\Bbag(3)/8$, which reproduces Eq.~\eqref{eq:bagvalue}
exactly. This coefficient is then held fixed as $N_c$ changes, since
refitting the bag to the same transition temperature at every $N_c$ would
remove the zero-field prediction.

The pure-glue limit of this family is $T_g=(b_g/a)^{1/4} \approx 196$ MeV, which
should be regarded as an extrapolation of the ideal model rather than as a determination of the Yang--Mills scale.

Before evaluating the curvatures, one must also specify how the charges
behave with $N_c$.
Reference~\cite{Fraga:2012ev} uses $((N_c-1)/N_c,-1/N_c)$ for a pair of flavors,
in which case only the first charge survives at leading order. We implement
this choice in addition to fixed physical charges. Both give the same
charges at $N_c=3$, and the pion charge remains equal to one. For
degenerate quarks the leading sum of squared charges changes from $5/9$ to
1, multiplying both $k_{1F}$ by $9/5$, while the subleading response
coefficient becomes $j_F=-2r_F^u-h_F$, where $r_F^u$ is the response,
thermal plus vacuum, of the first flavor with unit charge and $h_F$ is the
pion response.

\begin{figure*}[t]
\includegraphics[width=\textwidth]{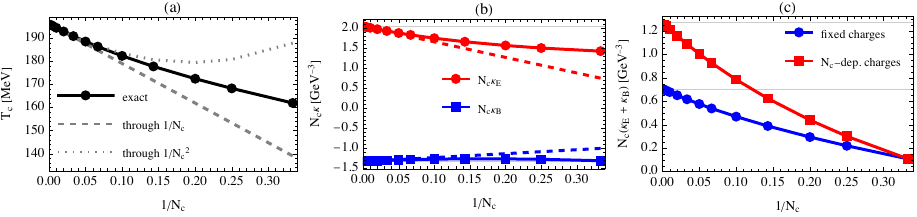}
\caption{Large-$N_c$ behavior with two quarks with mass $m_f=5$~MeV, pions with mass $m_\pi=135$~MeV,
$\mu_0=m_\pi$ and $\Bbag(N_c)/(N_c^2-1)$ fixed. Left: exact zero-field crossing and its first
two asymptotic approximations. Middle: scaled electric and magnetic
curvatures, where dashed lines retain $k_1+k_2/N_c$ and horizontal lines
mark $k_1$. Right: scaled sum for fixed charges and for the charge
assignment of Ref.~\cite{Fraga:2012ev} (labeled $N_c$-dep). The two trajectories coincide at $N_c=3$,
but have different limits.}
\label{fig:nc}
\end{figure*}

With fixed charges $(2/3,-1/3)$ we find
\begin{gather}
 t_1=-172~\mathrm{MeV},\quad t_2=440~\mathrm{MeV},\nonumber\\
 (k_{1E},k_{1B})=(2.04,-1.33)~\mathrm{GeV}^{-3},\nonumber\\
 (k_{2E},k_{2B})=(-3.88,0.98)~\mathrm{GeV}^{-3}.
 \label{eq:ncnumbers}
\end{gather}
The exact roots of the gas model are compared with the asymptotic
expansion in Fig.~\ref{fig:nc}, where ``exact'' refers only to the
specified free-gas pressure and quadratic response.

As is clear from the figure, the expansion is not quantitatively
controlled at $N_c=3$. Its first two approximations give $\Tc=139$ and 188 MeV, to be compared with the exact value of 162 MeV, and
$\kappa_E=0.679$ and $0.248~\mathrm{GeV}^{-3}$, to be compared with Eq.~\eqref{eq:referencek}. For the magnetic curvature one obtains $-0.444$ and $-0.335~\mathrm{GeV}^{-3}$. Keeping the next term, therefore, does not systematically improve the result at the physical number of colors, even though the leading magnetic term happens to be close to the exact value. At large $N_c$, on the other hand, independent numerical tests recover the expected powers of the truncation remainder.

\subsection{The susceptibility sum and perturbative QCD}

At leading order, the Lorentz-invariant relative term cancels from the sum,
\begin{equation}
 \lim_{N_c\to\infty}N_c(\kappa_E+\kappa_B)
 =-\frac{\sum_f Q_f^2(\xi_f+\chi_f)}{a_1},
 \label{eq:ncsum}
\end{equation}
which is positive for free massive quarks. In the fixed-charge baseline this limit is $0.707~\mathrm{GeV}^{-3}$, whereas the scaled value at $N_c=3$ is only $0.107~\mathrm{GeV}^{-3}$. As one might expect, the small difference between two sizable curvatures is especially sensitive to subleading contributions.

For comparison, the massless high-temperature result at $\mu_B=0$ follows directly from Eq.~(4.1) of Ref.~\cite{Ferreira:2026qqu}:
\begin{equation}
 \xi+\chi=-\frac{N_c\sum_fQ_f^2}{12\pi^2}
 \left[1+\frac{\alpha_s C_F}{\pi}
 \left(4\ln2-\frac54\right)+\OO(\alpha_s^2)\right].
 \label{eq:pqcd}
\end{equation}
Notice that the explicit logarithms of the electromagnetic subtraction
scale cancel in this combination, as does the displayed leading soft
$\OO(\alpha_s^{3/2})$ contribution. The hard correction, on the other hand,
survives at large $N_c$, since
\begin{equation}
 \frac{\alpha_s C_F}{\pi}
 =\frac{\lambda}{8\pi^2}(1-N_c^{-2}).
 \label{eq:lambda}
\end{equation}
This is a direct consequence of the perturbative
result, and provides a
high-temperature check on the combined observable. In the 't Hooft limit,
higher orders must be organized in powers of $\lambda$, so that the
remainder in Eq.~\eqref{eq:pqcd} is not uniformly suppressed as $1/N_c^2$.

Of course, Eq.~\eqref{eq:pqcd} cannot be consistently used to improve the
curvature near the transition while leaving the pressure and the entropy
jump at ideal-gas order. Going beyond the present realization requires the
dressed response and the equation of state in the same scheme and regime,
and large color number alone does not make such a calculation perturbative.

\subsection{Possible lattice tests}
\label{sec:lattice}

Our analysis suggests that the most direct lattice observables would be the
susceptibility sum and the dependence of the two curvatures on the number
of colors. At a resolved first-order transition, with latent heat
$L=\Tc\Delta s$, coexistence implies
\begin{equation}
 -\frac{L\kappa_E}{\Tc}=\Delta\xi,\qquad
 -\frac{L\kappa_B}{\Tc}=\Delta\chi,
 \label{eq:lattice}
\end{equation}
which should be regarded as consistency relations rather than as
additional independent predictions. A study at fixed $N_f$, with matched
masses and specified charges, could then compare $N_cT_g^3\kappa_E$,
$N_cT_g^3\kappa_B$ and their sum as functions of $1/N_c$. Away from the transition, the ratio $(\xi+\chi)/(N_c\sum_fQ_f^2)$ provides a direct test of the thermal response and of its perturbative limit. Given the smallness
of this combination, correlated uncertainties in $\xi$ and $\chi$ must be
taken into account. At the physical-color crossover, instead, one must
compare the response with a specified pseudocritical criterion, without
assigning it a fictitious latent heat.

\section{Conclusions}
\label{sec:conclusions}

We have investigated the effect of weak electric and magnetic backgrounds on the thermal deconfinement
transition within a simple phenomenological description: a gas of pions for the hadronic sector and
the MIT bag model for the quark–gluon plasma. Besides the thermal response of each phase, we included the one-loop electromagnetic vacuum contributions of quarks and charged pions, renormalized at a common scale chosen so that the electric charge is that of the confined vacuum. Since the charged particles differ in the two phases, these contributions do not cancel, and they fix the relative vacuum term between the phases in terms of the particle masses. 

Adopting the static electric prescription employed in recent lattice simulations, we find that, {\it without any fitting of the electromagnetic response}, the critical temperature increases with the electric field and decreases with the
magnetic field, with an electric curvature of the same size as the lattice result and a magnetic one
between 2.5 and 5 times larger in magnitude than lattice estimates. The (correct) signs of the curvatures 
persist under all physical values of masses, {\it without any field-dependent fit}.

We also discussed how these results depend on the renormalization scale, and show how a large-$N_c$ expansion
organizes both curvatures. In particular, their sum is independent of the vacuum terms and can be directly compared to recent perturbative QCD results for the electromagnetic susceptibilities.

A crucial ingredient is the electromagnetic vacuum contribution of each phase, which we have computed at one loop with a common renormalization scale, chosen such that $e$ is the charge of the confined vacuum. Since the charged particles differ in the two phases, a finite relative vacuum term remains, and it reduces both curvatures by a factor of about four. The resulting electric curvature, $\kappa_E\simeq0.47~\mathrm{GeV}^{-3}$, is of the same size as the lattice value, while the magnetic one, $\kappa_B\simeq-0.44~\mathrm{GeV}^{-3}$, is between 2.5 and 5 times larger in magnitude than lattice estimates, as mentioned above. Of course, given the first-order nature of the bag transition and the dependence on the renormalization scale, this agreement should be regarded as semi-quantitative.

The magnetic sign follows from a positive entropy jump, from the
diamagnetism of the pions, and from the paramagnetism of the quark phase,
which, once the vacuum term is included, survives for temperatures above
about $0.57\,\mu_0$. The electric sign, in turn, follows from the fact that the sum $\xi+\chi$ of the quark phase is fixed by its thermal part, so that a paramagnetic quark phase has a large negative electric susceptibility. For heavy quarks the vacuum term dominates and increases both curvatures. The sum of the curvatures does not depend on the vacuum terms at all, although it is neither a universal order parameter nor protected against general medium corrections.

The large-$N_c$ construction organizes both curvatures at order $1/N_c$ and the pion response at order $1/N_c^2$. While the $1/N_c$ suppression and the relation between field response and
transition shift are general, the individual signs require dynamical input for the \emph{net} responses. The stronger constraint follows when
$r_E+r_B<0$. In this case, a negative magnetic curvature $\kappa_B$ in the large $N_c$ regime implies a positive electric curvature with larger magnitude, $\kappa_E>|\kappa_B|$. An explicit realization in the bag model illustrates the counting, but also an example of poor quantitative convergence of the truncated expansion at $N_c=3$, especially for observables with large cancellations. The qualitative behavior, however, remains the same down to the physical number of colors, $N_c=3$. The susceptibility sum connects this framework to the known hard and soft perturbative results, and provides a useful observable for lattice simulations at several numbers of colors. 

Although our description of the deconfinement transition in the presence of external electromagnetic fields is admittedly very simple, it provides the relevant ingredients for a semi-quantitative description of the behavior of the critical temperature as a function of $E$ and $B$. As discussed previously~\cite{Fraga:2012fs}, it seems that its focus on incorporating confinement (or the competition between two phases with different degrees of freedom) makes it suited to describe the behavior of $T_c$ as a function of external control parameters. In fact, we have previously achieved a successful description of the behavior of the critical temperature as a function of the pion mass and isospin chemical potential, as compared to lattice data, where chiral models failed even qualitatively~\cite{Fraga:2008be,Palhares:2008px}.

\begin{acknowledgments}
This work was partially supported by INCT-FNA (Process No. 464898/2014-5), CAPES (Finance Code 001), CNPq, and FAPERJ.
\end{acknowledgments}

\bibliography{elebag}

\end{document}